\documentclass[]{JHEP3}

\newcommand{\be}{\begin{equation}}
\newcommand{\ee}{\end{equation}}
\newcommand{\bea}{\begin{eqnarray}}   
\newcommand{\eea}{\end{eqnarray}}
\newcommand{\demu}{\partial_{\mu}}

\newcommand{\de}{\partial}
\newcommand{\bear}{\begin{array}{l}}
\newcommand{\eear}{\end{array}}

\newcommand{\ldl}{\Lambda \partial_{\Lambda}}
\newcommand{\inte}{\! \int \!\!}
\newcommand{\ie}{{\it i.e.}\ }
\newcommand{\eg}{{\it e.g.}\ }
\newcommand{\etal}{{\it et al.}}

\newcommand{\etc}{{\it etc.}\ }

\newcommand{\one}{\hbox{1\kern-2mm l}}

\def\lam{\lambda}
\def\Lam{\Lambda}
\def\de{\partial}

\def\0{\vec{0}}

\def\sig3{\sigma_3}

\def\ker#1{\cdot #1 \cdot}

\def\e{{\rm e}}

\def\hS{\hat{S}}

\def\one{\hbox{1\kern-.8mm l}}

\def\eq#1{eq.~(\ref{#1})}
\def\ceq#1{Eq.~(\ref{#1})}

\def\s#1#2#3{S^{(#1)}_{#2} (#3)}
\def\phi{\varphi}
\def\tp{\tilde{\phi}}
\def\tS{\tilde{S}}
\def\thS{\tilde{\hS}}
\def\tg{\tilde{\gamma}}

\title{ Exact scheme independence at one loop }

\author{Stefano Arnone, Antonio Gatti
and Tim R. Morris\\ 
Department of Physics and Astronomy,
University of Southampton\\
\hspace{.045em} Highfield, Southampton SO17 1BJ, U.K.\\

E-mails: \email{sa@hep.phys.soton.ac.uk, gatti@hep.phys.soton.ac.uk, 
T.R.Morris@soton.ac.uk}}  

\maketitle

\preprint{SHEP 02-01}
\abstract{The requirement that the quantum partition function be 
invariant under a renormalization group transformation results in a 
wide class of exact renormalization group
equations, differing in the form of the kernel. Physical quantities
should not be sensitive to the particular choice of the kernel. We
demonstrate this scheme independence in four dimensional scalar 
field theory by showing that, even with a general kernel, the one-loop 
beta function may be expressed only in terms of the effective action 
vertices, and thus, under very general conditions, 
the universal result is recovered.}

\begin{document}
\section{Introduction}

The non-perturbative meaning of renormalization, as understood
by Wilson, is formulated most directly in the continuum in terms
of the exact Renormalization Group (RG) \cite{Wil}. Moreover, the
fact that solutions $S$ of the corresponding flow equations
can then be found directly in terms of renormalized quantities, that
all physics (\eg Green functions) can be extracted from this
Wilsonian effective action $S$, and that renormalizability is 
trivially preserved in almost any approximation \cite{morig,rev}, turns these 
ideas into a powerful framework for considering non-perturbative
analytic approximations (see for example refs.\ \cite{Wil}--\cite{general},
\cite{We}, \cite{Mo1}--\cite{Mo}, \cite{Tighe}). 

In the past a number of different versions and ways of deriving
the exact RG have been proposed \cite{Wil}--\cite{Mo1},
but in fact the resulting flow 
equations may readily be shown to be equivalent under changes of 
variables \cite{morig,rev,truncam,sumi,LM}. 

Recently, far more general versions
of the exact RG have been considered, dependent on the choice of
a functional $\Psi$, known as the ``kernel'' of the exact RG \cite{Mo1,LM}.
This elucidates the structure behind all
forms of exact RG \cite{LM}. In particular, each exact
RG is associated with a $\Psi$, that induces a
reparametrization (field redefinition) along the flow, and acts
as a connection between the theory space of actions at different
scales $\Lambda$. As a result, local to some generic point $\Lambda$
on the flow, all the exact RGs, including these generalised ones,
may be shown to be just reparametrizations of each other. 
When this reparametrization can be extended globally, the result is
an immediate proof of scheme
independence for physical observables. Indeed computations of physical
quantities then differ only through some field reparametrization.
One practical example
is an explicit field redefinition that interpolates between 
results computed using different choices of cutoff function 
$c(p^2/\Lambda^2)$ \cite{LM}.

Even more dramatic than this however, is the use of this freedom to
adapt the exact RG to certain forms of approximation or special
physical problems \cite{LM}. In particular, recently there has been substantial
progress in adapting these ideas to gauge theory. 
It turns out that not only can one introduce an effective cutoff $\Lambda$
in a way that does not break the gauge invariance \cite{gareg} but 
careful choices of $\Psi$ allow the gauge invariance to be
preserved manifestly (\ie not even gauge fixed) along the flow and
in the solutions $S$ \cite{Mo1,Mo,ant}.

Nevertheless, the analysis presented in ref.\ \cite{LM} leaves open a 
number of practical questions. Although at general points of the flow,
all exact RGs are locally equivalent,
obstructions can arise to full (global) equivalence, 
on the one hand from differences in the global structure of fixed points 
deduced from the two flows,
and on the other from the non-existence at special points of an
inverse in an implied change of variables (from $S$ to $\Lambda$) 
\cite{LM}. However it is difficult to see in general how one can 
determine when such obstructions do exist, given that 
in practice one has to make approximations 
in order solve these theories. Furthermore, computations within a 
generalised exact RG, such as the type being used for gauge theory 
\cite{Mo1,Mo},
generate many more terms, whose link to standard text-book methods of
doing quantum field theory seem quite obscure. Computations of physical
quantities (such as the large $N$ $SU(N)$ Yang-Mills one-loop
$\beta$ function) fall out as the universal expected result
but only after a complicated calculation and apparently by magic 
\cite{Mo1,Mo}.

This paper addresses the above problems within a sufficiently simple
and bounded context: the one-loop $\beta$ function of four dimensional 
(one component) scalar field theory. We will see that even for a very
general form of $\Psi$ (one involving a general `seed' action $\hS$),
the correct universal result is obtained. To our knowledge, this is the
first concrete
test of such scheme independence beyond testing for cutoff function
independence. The only requirements we have
to impose on $\hS$ to recover the universal result,
are some very weak and general requirements which are
necessary in any case to ensure that the
Wilsonian action $S$ makes sense. To this level then, all such exact
RGs are equivalent and merely parametrise changes of scheme. 

From a practical point of view, we find that a major step
in understanding and solving
these equations comes from using the flow equations to exchange 
elements of $\hS$ (more generally $\Psi$) in favour of vertices from
the Wilsonian action $S$. 
Some expressions for the quantum corrections in $S$ 
then turn out to simplify
dramatically and result in scheme `covariant' expressions with 
straightforward physical interpretations. The one-loop $\beta$
function, $\beta_1$, 
is one example where this covariance then implies even 
{invariance} under changes of scheme. Indeed, we find very directly
in this way, the same answer for $\beta_1$
independent of the details of the cutoff function
$c$ and the seed action $\hS$.

The plan of the paper is as follows. In sec.\ 2, we review briefly
one version of the exact RG, Polchinski's equation \cite{Po}, use this
a basis to explain the more general exact RGs, and introduce our
concrete set of such things, involving $\hS$. In sec.\ 3, we perform
the one-loop calculation using this general form of exact RG. Finally,
in sec.\ 4, we summarise and draw our conclusions.

\section{From the Polchinski equation to general exact RGs}

We will consider a massless scalar field theory in four Euclidean 
dimensions, with a momentum space cutoff.
The theory is defined at some ultra-violet scale, $\Lam_0$, by giving the 
quantum partition function, 
\be
{\cal Z}_{\Lam_0}= \int\!\! {\cal D}[\varphi] \, \e^{-S_{\Lam_0}[\varphi]} . 
\ee
The action consists of the kinetic term, regularised by the introduction of
a cutoff function, and the characteristic self-interaction term,
\be \label{bareaction}
S_{\Lam_0}[\varphi] = {1\over 2}\inte {d^4 p
\over (2\pi)^4} \, p^2 c^{-1}({\textstyle {p^2\over \Lambda_0^2}}) \, \varphi^2 + {\lam_0 
\over 4!} \inte d^4 x \, \varphi^4 + {m_0^2 
\over 2} \inte d^4 x \, \varphi^2.
\ee 
Here $c(p^2/\Lambda^2)>0$ is a {\it smooth}, \ie infinitely differentiable,
ultra-violet cutoff profile.
The cutoff, which modifies propagators $1/p^2$ to $c/p^2$, 
satisfies $c(0)=1$ so that low energies are unaltered,
and $c(p^2/\Lambda^2)\to0$ as $p^2/\Lambda^2\to\infty$
sufficiently fast that all Feynman diagrams are ultra-violet regulated.

The RG transformation amounts to changing the cutoff from
$\Lam_0$ to $\Lam << \Lam_0$ and compensating for such change by replacing
the action with a more complicated effective action, 
{\it a.k.a.} the Wilsonian action \cite{Wil}, 
\be
\label{wilsact}
S[\varphi] = {1\over 2} \inte {d^4 p
\over (2\pi)^4} \, p^2 c^{-1}(p^2/\Lambda^2) \, \varphi^2 +
S^{int}[\varphi;\Lam].    
\ee
Demanding that physics be invariant under the RG transformation
results in a 
functional differential equation for the effective interaction \cite{Po},    
\be \label{sintfleq}
\ldl S^{int} = -{1\over\Lambda^2}
 {\delta S\over\delta\phi}^{int} \!\!\!\!\!\ker{c'}  
{\delta S\over\delta\phi}^{int}\!\!\! +{1\over\Lambda^2}{\delta 
\over\delta\phi} \ker{c'} {\delta S\over\delta\phi}^{int}\!\!\!\!\!\!.
\ee
In \eq{sintfleq}, prime denotes differentiation with respect to the 
function's argument (here $p^2/\Lambda^2$) and the following shorthand has 
been introduced: for two functions $f(x)$ and $g(y)$ and a 
momentum space kernel $W(p^2/\Lambda^2)$, 
with $\Lam$ being the effective cutoff, 
\be
f \ker{W} g =
\int\!\!\!\!\int\!\!d^4\!x\,d^4\!y\
f(x)\, W_{x y}\,g(y),
\ee
where $W_{x y} = \inte {d^4 p \over (2\pi)^4} \, W(p^2/\Lambda^2) {\rm
e}^{i p \cdot (x-y)}$.

Note that the regularised kinetic term in \eq{wilsact}
may be written as 
\be
\label{ke}
{1\over 2}\,\demu \varphi \ker{c^{-1}} \demu\varphi. 
\ee
This will be referred to as the seed action and denoted 
by $\hS$. In terms of the total effective action, $S[\varphi]=\hS+S^{int}$, 
and $\Sigma \doteq S-2\hS$, the exact RG equation reads
\be \label{pofleq}
\ldl S = -{1\over\Lambda^2}
 {\delta S\over\delta\varphi} \ker{c'}  
{\delta \Sigma \over\delta\varphi} +{1\over\Lambda^2}{\delta 
\over\delta\varphi} \ker{c'} {\delta \Sigma \over\delta\varphi}
\ee
(up to a vacuum energy term that was discarded in~\cite{Po}). 

The invariance of the partition function is manifest if \eq{pofleq} is
recast as
\be
\label{measurefl}
\ldl \e^{-S} =
-{1\over\Lambda^2}{\delta\over\delta\varphi}\cdot c' \cdot \left(
{\delta\Sigma \over\delta\varphi}\,\e^{-S}\right),
\ee
\ie the infinitesimal RG transformation results in a change in the
integrand which is a total functional derivative.

Incidentally, this estabilishes a rather counterintuitive result, that
integrating out degrees of freedom is just equivalent to redefining the
fields in the theory \cite{Mo,LM}. In the present case, the change in
the partition function may be shown to correspond to the change of
variables $\varphi \rightarrow \varphi + \delta \Lam \, \Psi$, with the
``kernel'' $\Psi = -{1\over \Lam^3} \, c' \, {\delta \Sigma
\over\delta\varphi}$ that appears in \eq{measurefl} \cite{LM}.

Different forms of exact RG equations correspond to choosing different
kernels $\Psi$. There is a tremendous amount of freedom in this choice:
more general seed actions $\hS$ can be considered,
$c'$ can be augmented by further terms including interactions (analogous
to the ``wines'' in refs.\ \cite{Mo1,Mo}), higher functional derivatives
and/or more complex dependences on the effective action $S$ can
be included \cite{LM}. 

Intuitively, we should expect that there is a wide freedom in the choice of 
kernel, just as there is a great deal of freedom in choosing the form
of a blocking transformation in the condensed matter or lattice 
realisation of the Wilsonian renormalization group \cite{Wil}.
The flow equation \eq{sintfleq} is distinguished only by its relative
simplicity (related to incorporating the cutoff only in the kinetic
term). Nevertheless, physical quantities should turn out to be universal,
 \ie independent of these choices.
It can be shown that local to generic points 
$\Lambda$, all of these choices are actually related to each other by 
field redefinitions \cite{LM}. If no obstructions exist to extending
this for all $\Lambda$, then universality follows immediately.
However obstructions can arise from the inability to 
invert these maps and from global restrictions (in particular arising
from the structure of fixed points) \cite{LM}. 
In this paper we investigate these issues at a completely concrete
level, computing the one-loop $\beta$ function in scalar field
theory with a generalised $\Psi$. With so much freedom we 
have to restrict it to be able to be concrete;
we choose to consider a general seed action $\hS$ of a form that we now
outline.

As shown above, Polchinski's equation comes from setting the seed action
equal to the effective kinetic term in the Wilsonian effective action 
\eq{wilsact}. As we will show in the 
next section, if we are to reproduce at the classical level
the form of the effective kinetic term in \eq{wilsact}, 
the bilinear term in $\hS$
must continue to be equal to \eq{ke}.\footnote{It
is not necessary that the classical kinetic term take the
form in \eq{wilsact}. We choose to require it purely for convenience.}
Furthermore, we choose to leave the $\phi\leftrightarrow-\phi$ symmetry 
alone, by requiring that $\hS$ is even under this symmetry. We are left 
with a generalised exact renormalization group parametrised by the 
infinite set of seed action $n$-point vertices, $n=4,6,8,\cdots$. We will
leave each of these vertices as completely unspecified functions of
their momenta except for the requirement that {\it the vertices be infinitely 
differentiable and lead to convergent momentum integrals}.
(The first condition ensures that no spurious infrared 
singularities are introduced and that all effective vertices can be
Taylor expanded in their momenta to any order \cite{Mo1,rev}.
The second condition is necessary for the flow equation to make sense
at the quantum level and also ensures that the flow actually corresponds
to integrating out modes \cite{Mo,LM}.)

We are therefore incorporating in the momentum dependence of
{\sl each} of the seed action $n$-point vertices, $n=2,4,6,\cdots$, 
an infinite number of parameters.
Of course these infinite number of vertices, each with an infinite number
of parameters, then appear in the effective action $S$ as a consequence
of the flow equation. Remarkably, we can still compute the one-loop 
$\beta$ function. Moreover, as we will see in the next section, we can
invert the flow equation by expressing $\hS$ vertices in terms of
$S$, and in this way demonstrate explicitly that the result is universal
- {\it viz.} independent of the choice of $c$ and $\hS$.
It follows that, at least in this case, the {\sl only} requirements that
we need to impose on $\hS$ in order to ensure that these generalised flow 
equations continue to describe the same physics are those italicized in 
the previous paragraph.

\section{One-loop beta function with general seed action}

Physical quantities must be universal, \ie independent of the
renormalization scheme. In particular, they should not be sensitive to 
the particular choice of
the RG kernel, \eg on the form of the cutoff function or the expression for the
seed action. 
We aim to calculate one of those, the one-loop contribution to the $\beta$
function, while keeping as general a seed action as possible. 
As we will see, an elegant, clear cut way of achieving such a result is to 
make use of the equations of motion for the effective couplings in order 
to get rid of the seed action vertices. 

As usual, universal results are obtained only after the imposition of 
appropriate renormalization conditions which allow us to define what we
mean by the physical (more generally renormalized) coupling and field. 
(The renormalized mass must also be defined and is here set to zero 
implicitly by ensuring that the only
scale that appears is $\Lambda$.) 

We write the vertices of $S$ as
\be
\s{2n}{}{\vec{p};\Lam}\equiv\s{2n}{}{p_1,p_2,\cdots,p_{2n};\Lam}
\doteq (2\pi)^{8n} 
{\delta^{2n} S\over\delta\phi(p_1)\delta\phi(p_2)\cdots\delta\phi(p_{2n})},
\ee
(and similarly for the vertices of $\hS$). In common with earlier works 
\cite{Po,Bo}, we define the renormalized four-point coupling $\lambda$ by 
the effective action's four-point vertex evaluated at zero momenta:
$\lambda(\Lambda)=\s{4}{}{\vec{0};\Lam}$. This makes sense once we 
express quantities in terms of the renormalized field,
defined (as usual) to bring the kinetic term
into canonical form $\s{2}{}{p,-p;\Lam}= \s{2}{}{0,0;\Lam} + p^2
+O(p^4/\Lam^2)$.  
The flow equation can then be taken to be of the form \cite{sumi,ball}:
\be \label{gammafleq}
\ldl S - {\gamma\over2}\ \phi\!\cdot\!{\delta S\over \delta \phi}
   = -{1\over\Lambda^2}
 {\delta S\over\delta\varphi} \ker{c'}  
{\delta \Sigma \over\delta\varphi} +{1\over\Lambda^2}{\delta 
\over\delta\varphi} \ker{c'} {\delta \Sigma \over\delta\varphi}.
\ee
We have used the short hand
\be
\nonumber
\phi\!\cdot\!{\delta S\over \delta \phi} \doteq
\inte {d^4 p\over (2\pi)^4} \, \phi(p) 
{\delta S\over \delta \phi(p)}
\ee
and as usual the anomalous dimension $\gamma={1\over Z}\ldl Z$,
where $Z$ is the wavefunction renormalization. As emphasised
in refs.\ \cite{Mo1,LM}, although
\eq{gammafleq} is not the result of changing variables 
$\phi\mapsto \phi\sqrt{Z}$ in \eq{pofleq}, it is still a perfectly
valid flow equation and a more appropriate starting point when
wavefunction renormalization has to be taken into 
account. This is in fact a small example of the immense freedom we have
in defining the flow equation.
(The new term on the left hand side arises from replacing 
$\partial_\Lam |_\phi$ with a partial derivative at constant renormalized
field, but in order to produce the right hand side, and in order
to reproduce the same $\hS$, we need to start with the alternative 
cutoff function $cZ$ in eqs.\ (\ref{bareaction}) -- (\ref{pofleq}).
Alternatively, for the purposes of computing the $\beta$ function, we 
could have simply taken account of the wavefunction renormalization 
afterwards as in ref.\ \cite{Tighe}.)

We now rescale the field $\varphi$ to 
\be
\label{rescale}
\varphi =\frac{1}{\sqrt{\lambda}}\, {\tp},
\ee 
so as to put the coupling constant in
front of the action. This ensures the expansion in the coupling constant 
coincides with the one in $\hbar$, the actual expansion parameter being
just $\lam \hbar$. The resulting expansion is more elegant, 
being no longer tied at
the same time to the order of expansion of the field $\phi$. It is also
analogous to the treatment pursued for gauge theory in refs.\ \cite{Mo1,Mo}
(where gauge invariance introduces further simplications in particular
forcing $\gamma=0$ for the new gauge field). The following analysis thus
furnishes a demonstration that these ideas also work within scalar
field theory.  

The bare action (\ref{bareaction}) rescales as 
\be
S_{\Lam_0}[\varphi] = 
{1\over \lam(\Lam_0)} \left[{1\over 2} \inte {d^4 p
\over (2\pi)^4} \, p^2 c^{-1}({\textstyle {p^2\over \Lambda_0^2}}) \, 
{\tp}^{2} + {1 
\over 4!} \inte d^4 x \, {\tp}^{4} + {m_0^2 
\over 2} \inte d^4 x \, {\tp}^{2} \right] \doteq {1\over \lam(\Lam_0)}
{\tS}_{\Lam_0}[{\tp}]. 
\ee

Defining the ``rescaled'' effective and seed actions as $S[\varphi] = 
{1\over \lam} \tS[\tp], 
\hS[\varphi] = {1\over \lam}\thS
[\tp]$, and absorbing the change to $\partial_\Lam|_{\tp}$
in a change to $\tg$,
the flow equation (\ref{pofleq}) reads
\be
\label{tildefleq}
\ldl \left({1\over \lam} \tS\right)
- {\tg\over2\lam}\ \tp\!\cdot\!{\delta \tS\over \delta \tp}
=-\frac{1}{\lambda \Lam^2}\frac{\delta
(\tS-2\thS)}{\delta\tp} 
\ker{c'} \frac{\delta \tS}{\delta
\tp}+\frac{1}{\Lam^2}\frac{\delta}{\delta
\tp}\ker{c'}\frac{\delta(\tS-2\thS)}{\delta\tp}. 
\ee
Expanding the action, the beta function $\beta(\Lam) = \ldl \lam$ 
and anomalous dimension, in
powers of the coupling constant:
\bea \nonumber
\tS[\tp]&=&\tS_0+\lambda \tS_1+\lambda^2
\tS_2+\cdots,\\ \nonumber
\beta(\Lam)&=&\beta_1\lambda^2+\beta_2\lambda^3+\cdots,\\ \nonumber
\tg(\Lam) &=&\tg_1\lam+\tg_2\lam^2+\cdots
\eea
yields the loopwise expansion of the flow equation\footnote{In order to 
simplify the notation, the tildes will be removed from now on.}
\bea
&&\Lam\de_{\Lam}S_0=-\frac{1}{\Lam^2}\frac{\delta S_0}{\delta\varphi}\cdot
c'\cdot\frac{\delta (S_0-2\hat{S})}{\delta\varphi}\label{scalartree},\\
&&\Lam\de_{\Lam}S_1-\beta_1 S_0-{\gamma_1\over2}\ 
\phi\!\cdot\!{\delta S_0\over\delta\phi}
=\nonumber \\
&&\phantom{\Lam\de_{\Lam}S_1-\beta_1 S_0}
-\frac{2}{\Lam^2}\frac{\delta S_1}{\delta\varphi}\cdot
c'\cdot\frac{\delta(S_0-\hat{S})}{\delta\varphi}+\frac{1}{\Lam^2}\frac{
\delta}{\delta\varphi}\cdot
c'\cdot\frac{\delta(S_0-2\hat{S})}{\delta\varphi}\label{scalar1loop},
\\ \nonumber
\eea
\etc
$\gamma_1$ and $\beta_1$ may now be extracted directly from 
\eq{scalar1loop},  as specialised to the two-point and four-point 
effective couplings, $\s{2}{}{\vec{p};\Lam}$ and $\s{4}{}{\vec{p};\Lam}$ 
respectively, once the renormalization conditions have been taken
into account. 

We impose the wavefunction renormalization condition
in the new variables:
\be
\label{r2}
\s{2}{}{p,-p;\Lam} = \s{2}{}{0,0;\Lam}+ p^2 +O(p^4/\Lam^2).
\ee
Bearing in mind that the coupling constant has been scaled 
out, we impose the condition
\be
\label{r4}
\s{4}{}{\vec{0};\Lam} = 1.
\ee
Both conditions \eq{r2} and \eq{r4} are already saturated at tree level. 
(To see this it is sufficient to note that, since the theory is massless, 
the only scale involved is $\Lam$.
Since $S^{(4)}_0$ is dimensionless 
it must be a constant at null momenta, thus  
$\s{4}{0}{\vec{0}; \Lam} = \s{4}{0}{\vec{0}; 
\Lam_0} = 1$. Similar arguments apply to $S^{(2)}_0$.)
Hence the renormalization condition implies that we must have 
no quantum corrections to the four-point vertex at $\vec{p} =
\vec{0}$, or to the $O(p^2)$ part of the two-point vertex,
\ie
\be
\s{4}{n}{\vec{0}; \Lam} = 0 \quad{\rm and}\quad
\left.\s{2}{n}{p,-p;\Lam}\right|_{p^2} =0\qquad \forall n \geq 1,
\ee
where the notation $|_{p^2}$ means that one should take the coefficient 
of $p^2$ in the series expansion in $p$.
The flow equations for these special parts of the quantum corrections
thus greatly simplify, 
reducing to algebraic equations which then determine the $\beta_i$
and $\gamma_i$. In particular, from the flow of $S^{(4)}_1$
at null momenta:\footnote{Here and
later we suppress the $\Lambda$ dependence of the $S$ and $\hS$ vertices.} 
\be\label{beta1}
\beta_1+2\gamma_1=\frac{8c'_0}{\Lam^2}\Big[1-\hat{S}^{(4)}(\vec{0})\Big] 
\s{2}{1}{0}-\frac{1}{\Lam^2}\int_q c'({\textstyle {q^2 \over
\Lam^2}})\Big[S^{(6)}_0-2\hat{S}^{(6)}\Big](\vec{0},q,-q), 
\ee 
where $c'_0 = c'(0)$ and $\int_q \doteq \int {d^4 q \over (2 \pi)^4}$,
and from the flow of $S^{(2)}_1$ expanded to $O(p^2)$:
\be\label{gamma1}
\beta_1+\gamma_1= - \frac{1}{\Lam^2}\int_q c'({\textstyle {q^2 \over
\Lam^2}})
\Big[ \left.S^{(4)}_0-2 \hS^{(4)}\Big] (p,-p,q,-q)\right|_{p^2}.
\ee
Note that contrary to the standard text book derivation our one-loop
anomalous dimension is not zero, picking up a contribution from the
general field reparametrization \cite{LM} induced by higher point terms in 
$\hS$ and a contribution $-\beta_1$
due to the field rescaling \eq{rescale}.

In order to evaluate \eq{beta1}, we need to
calculate $\s{2}{1}{0}$ and $\s{6}{0}{\vec{0}, q, -q}$. We would also need
$\hS^{(4)}(\vec{0})$ and $\hS^{(6)}(\vec{0},q,-q)$, but we will see
that we can avoid using explicit expressions for them, and thus keep 
$\hS$ general, by using the equations of motion to express them in terms 
of the effective vertices $S^{(4)}_0$ and $S^{(6)}_0$. 

However, as explained in the previous section, our $\hS$ is not
completely arbitrary. Apart from some very general 
requirements on the differentiability and integrability of its vertices, 
for convenience we restrict $\hS$ to have only even-point vertices, as in 
fact already used in eqs.\ (\ref{beta1}) and (\ref{gamma1}), and constrain
its two-point vertex so that the two-point effective coupling keeps the
same functional dependence upon $\Lambda$ as the bare one (as in
\eq{wilsact}). This last condition reads 
\be \label{s20}
\s{2}{0}{p} = p^2 c^{-1} ({\textstyle {p^2\over \Lam^2}})
\ee
and from the two-point part of \eq{scalartree}, we immediately find
\be \label{shat20}
\hS^{(2)}(p) = p^2 c^{-1} ({\textstyle {p^2\over \Lam^2}}).
\ee
  
Let us start with the calculation of $\s{2}{1}{0}$. From \eq{scalar1loop},
its equation reads
\be
\ldl S_1^{(2)}(0) = \frac{1}{\Lam^2}\int_q c'({\textstyle {q^2 \over
\Lam^2}}) \Big[S^{(4)}_0 - 2 \hS^{(4)} \Big] (0,0,q,-q), \label{s12}
\ee
where eqs.\ (\ref{shat20}) and (\ref{s20}) have been already used to cancel 
out the classical terms.
Pursuing our strategy, we get rid of $\hS^{(4)}$ by making use of the
equation of motion for the four-point effective coupling at tree level
\be\label{sh40gen}
\ldl \s{4}{0}{\vec{p}} = {2\over \Lam^2} \sum_i
\frac{p_i^2 c'_{p_i}}{c_{p_i}} \hS^{(4)}(\vec{p}),
\ee
where $c_{p_i} \doteq c({p_i^2\over \Lam^2})$ and the invariance of
$\s{4}{0}{\vec{p}}$ under permutation of the $p_i$'s 
(which it has without loss of generality) has been utilised. 
Specialising the above equation to $\vec{p} = (0,0,q,-q)$, \eq{s12} becomes
\bea
\ldl \s{2}{1}{0} &=&\frac{1}{\Lam^2} \int_q
c'_q S_0^{(4)} (0,0,q,-q) -\int_q  \frac{c_q}{2 q^2} \ldl S_0^{(4)}
(0,0,q,-q)  \nonumber\\
&=& - \int_q {1\over 2 q^2} \ldl \Big( 
c_q \,  S_0^{(4)}(0,0,q,-q) \Big) \nonumber\\
&=&-\ldl \int_q \frac{c_q \,  \s{4}{0}{0,0,q,-q}}{2 q^2}.\label{eqs120}
\eea 
In 
the above, the derivative with respect to the cutoff may be taken after 
integrating over the loop momentum since the integral is regulated both
in the ultraviolet and in the infrared as a result of the properties of 
the effective couplings. \ceq{eqs120} may be now integrated
to give
\be \label{s2}
\s{2}{1}{0} = -\int_q \frac{c_q \, \s{4}{0}{0,0,q,-q}}{2 q^2},
\ee
with no integration constant since for a massless theory, 
there must be no other explicit scale in the theory apart from the 
effective cutoff. 

Let us now move on to the tree-level six-point function. From 
(\ref{scalartree}) we get
\bea
\ldl \s{6}{0}{\vec{0},q,-q} &=& \frac{4q^2}{\Lam^2}
\frac{c'_q}{c_q}\hat{S}^{(6)}(\vec{0},q,-q)\nonumber\\[3pt] 
&&-\frac{8c'_0}{\Lam^2}\Big[1-\hat{S}^{(4)}(\vec{0}) \Big]
\s{4}{0}{0,0,q,-q}
+\frac{8c'_0}{\Lam^2}\hat{S}^{(4)}(0,0,q,-q)\nonumber\\[3pt] 
&&-\frac{12}{\Lam^2}c'_q \, S^{(4)}_0(0,0,q,-q)
\Big[S^{(4)}_0-2\hat{S}^{(4)} \Big] (0,0,q,-q). \label{s6}
\eea
Using \eq{sh40gen}, and solving for $\hat{S}^{(6)}(\0,q,-q)$, 
\bea\label{sh6}
\hat{S}^{(6)}(\0,q,-q)&=&\frac{\Lam^2}{4q^2} \frac{c_q}{c'_q} \left\{ \ldl
\s{6}{0}{\0,q,-q)} +\frac{8 c'_0}{\Lam^2}\Big [1-\hat{S}^{(4)}(\0) \Big]
\s{4}{0}{0,0,q,-q} \right.\nonumber\\[3pt]
&&-2c'_0 \frac{c_q}{q^2c'_q} \ldl \s{4}{0}{0,0,q,-q}\nonumber\\[3pt]
&&\left.-\frac{6}{q^2} \s{4}{0}{0,0,q,-q} \ldl \Big[ c_q \,
\s{4}{0}{0,0,q,-q} \Big] \right\}.
\eea
We will see that
substituting eqs.\ (\ref{s2}) and (\ref{sh6}) into \eq{beta1}
will cause almost all the non universal terms to cancel
out. The remaining ones will disappear once $\gamma_1$ is substituted
using \eq{gamma1}, leaving just the precise form of
the one-loop beta function. 

Note that in \eq{sh6} and later, it appears
at first sight that we need to be able to take the inverse $1/c'_q$.
This would mean that in addition to the general restrictions on $\hS$
outlined earlier (and in the conclusions) we would also require that
$c'$ does not vanish at finite argument. In fact, we could arrange the
calculation more carefully so that $1/c'$ never appears, thus \eg here
we can recognize that only $c'_q \hat{S}^{(6)}(\0,q,-q)$ is needed for
\eq{beta1} and that from \eq{sh40gen}, $\ldl \s{4}{0}{0,0,q,-q}$ has 
a factor of $c'_q$. For clarity's sake, we will continue to write $1/c'$
in intermediate results and leave as an exercise for the reader to 
check that all such inverses can be eliminated.

Returning to the calculation in detail, 
the first term in (\ref{sh6}) and the $S^{(6)}_0$ term in
(\ref{beta1}) may be paired up into
\be \label{first}
\ldl \int_q \frac{c_q}{2 q^2} \, \s{6}{0}{\0,q,-q},
\ee
where again, due to the properties of the effective action vertices, the
order of the derivative and integral signs can be exchanged. Moreover, as
the integrand in \eq{first} is dimensionless, there cannot be any
dependence upon $\Lam$ 
after the momentum integral has been carried out, hence the result vanishes
identically!   
Also, the second term in (\ref{sh6}), when substituted into
(\ref{beta1}), exactly cancels the first term of the latter once
(\ref{s2}) is used. One is then left with
\bea\label{beta1tilde1}
\beta_1+2\gamma_1&=&-c'_0 \int_q \frac{c_q^2}{q^4
c'_q} \ldl \s{4}{0}{0,0,q,-q} - 3 \int_q \frac{c_q}{q^4}
\s{4}{0}{0,0,q,-q} \ldl \Big\{ c_q \, \s{4}{0}{0,0,q,-q} \Big\} 
\nonumber\\[3pt]
&=&-c'_0 \int_q \frac{c_q^2}{q^4
c'_q} \ldl \s{4}{0}{0,0,q,-q} - \frac{3}{2} \int_q \frac{1}{q^4}
\ldl \Big\{ c_q \, \s{4}{0}{0,0,q,-q} \Big\}^2.
\eea

In order to cancel out the first term in \eq{beta1tilde1}, the one-loop
contribution of the wave function renormalization coming from 
\eq{gamma1} must be taken into account. Again making use of \eq{sh40gen} 
to rid us of the hatted four-point coupling, 
\be\label{s4hp2}
\frac{1}{\Lam^2}\,\hS^{(4)}(p,-p,q,-q)\Big|_{p^2}=
\frac{c_q}{4 q^2
c'_q} \ldl S^{(4)}_0 (p,-p,q,-q) \Big|_{p^2}\!-\, c'_0 \left(
\frac{c_q}{2q^2c'_q}\right)^2\!\! \ldl \s{4}{0}{0,0,q,-q},
\ee
and substituting back in \eq{gamma1},
\be\label{bibita}
\beta_1+\gamma_1 =\frac{1}{2} \ldl \int_q
c_q \left. \s{4}{0}{p,-p,q,-q} \right|_{p^2}
-\frac{c'_0}{2}\int_q c'_q \left(\frac{c_q}{q^2c'_q}\right)^2 \ldl
\s{4}{0}{0,0,q,-q}. 
\ee
The first term on the right hand side of \eq{bibita} vanishes as it is a 
dimensionless UV and IR convergent integral, and therefore $\gamma_1$ 
takes the form
\be\label{zed1}
\gamma_1 = -\beta_1-\frac{c'_0}{2}\int_q c'_q
\left(\frac{c_q}{q^2c'_q}\right)^2 \ldl \s{4}{0}{0,0,q,-q}.
\ee
Finally, substituting (\ref{zed1}) in (\ref{beta1tilde1}) yields
\bea
\beta_1&=&\frac{3}{2} \int_q \frac{1}{q^4}\, \ldl\, \Big\{ c_q
\s{4}{0}{0,0,q,-q} \Big\}^2\label{betaf}\\[3pt]
&=&-\frac{3}{2}{\Omega_4\over(2\pi)^4} 
\int_0^{\infty}\!\!\! dq \, \de_q \left\{ c_q
\, \s{4}{0}{0,0,q,-q} \right\}^2 \nonumber\\[3pt]
&=&\frac{3}{16\pi^2},
\eea
which is the standard one-loop result \cite{PSGL}.\footnote{The term in 
braces depends only on $q^2/\Lam^2$. $\Omega_4$
is the four dimensional solid angle. The last line follows from the
convergence of the integral and
normalisation conditions $c(0)=1$ and (\ref{r4}). As far as independence
with respect to the choice of cutoff function is concerned, this is
standard.}
Note that in the top equation the $\Lam$ derivative
cannot be taken outside the integral, as this latter would not then
be properly regulated in 
the infrared. Moreover, had that been possible, it would have resulted in 
a vanishing beta function, as the integral is actually dimensionless.

\section{Summary and conclusions}

Starting with the generalised exact RG flow equation (\ref{gammafleq}),
we computed tree level two, four and six point vertices. At one-loop
we computed 
the effective mass $\s{2}{1}{0}$ and wavefunction renormalization 
$\gamma_1$. By combining all these with the flow of the one-loop 
four-point vertex at zero momentum, we arrived at eq.\ (\ref{betaf}),
which collapses to the expected universal result $\beta_1=3/(4\pi)^2$.

The flow equation we used 
differs from the Polchinski flow equation
(\ref{sintfleq}), equivalently \eq{pofleq}, because the seed action
$\hS$ is no longer set to be just the kinetic term (\ref{ke}), but
is generalised to include all arbitrary even higher-point vertices.
These are subject only to 
some very weak and generic restrictions which are recalled below.

In addition we added the anomalous dimension 
$\gamma$ term in \eq{gammafleq} to take account
of wavefunction renormalization. Normally this is needed only from 
two loops onwards, but the more general field reparametrisation induced
by the generalised $\hS$, means that a wavefunction renormalization $Z$ is
required for the effective action $S$ even at one loop. (The $\gamma$
term does not exactly follow from the flow equation (\ref{pofleq})
with cutoff $c$, but rather starts from one with cutoff $Zc$, but this
is more appropriate for cases where wavefunction renormalization is
involved.) As a final modification, we also scaled the coupling $\lambda$
out of the bare action, by rescaling the field. However the result 
(\ref{tildefleq}), is still an equivalent flow to (\ref{gammafleq}), 
since they are related by this simple change of variables. (In 
particular this means that the higher point vertices in $\hS$ are  those 
of \eq{gammafleq} multiplied by powers of $\lambda$.) Perturbative expansion
in $\lambda$ is now at its most elegant since it coincides with expanding
in $\hbar$ \ie the loop expansion. The structure also most closely 
coincides with the one used for gauge theory \cite{Mo1,Mo,ant}, so it acts 
as an independent test of this part of that framework.\footnote{Much less 
general unpublished tests were undertaken as preparation for the research 
in refs.\ \cite{Mo1,Mo}.} We then proceeded to compute the tree and one-loop corrections
exploiting the ability, within the exact RG, to derive 
directly the renormalized couplings and vertices (\ie without having to
refer back to an overall cutoff and bare action).

The effective action of the Polchinski flow equation,
\eq{sintfleq}, can be shown to be essentially the generating function
for connected diagrams in a field theory with infrared cutoff 
\cite{morig}. As a result the quantum corrections to this effective
action have a straightforward interpretation in terms of simple 
modifications of the usual
Feynman diagrams that follow from 
the original partition function \cite{morig}.
Since this direct link is obscured by the further field redefinitions
implied by the generalised $\hS$ \cite{LM}, we no longer expect the
diagrammatic interpretation of the quantum corrections to be quite so
simple. This expectation is indeed borne out by many of the equations
we presented, such as eqs.\ (\ref{beta1}), (\ref{gamma1}), (\ref{s6})
\etc Remarkably however, once $\hS$ vertices are
eliminated in favour of those of the Wilsonian effective action $S$,
two of our results do have such simple interpretations. One of these
is \eq{s2} which is the one-loop effective mass term required to ensure
that the theory is massless once all of the one-loop calculation is
completed (\ie once $\Lambda\to0$). One can see that it is nothing but
the usual tadpole term, formed from the classical effective four point
coupling $\s{4}{0}{\vec{q};\Lam}$ and the regularised propagator
$c_q/q^2$. Note that the result of the integral is {\sl not} universal:
it depends on the details of $c$, $S^{(4)}_0$ and thus $\hS$ \etc
\hspace{-.4em},
but the form it takes is invariant under scheme changes. It is therefore
in this sense, {\sl scheme covariant}. Another scheme covariant expression,
and again with a simple diagrammatic interpretation, is \eq{betaf}.
This is nothing but the standard one-loop diagrammatic result for 
the $\beta$ function, again written in terms of $c_q/q^2$ and the
effective $S^{(4)}_0$. As we saw, from here it is straightforward to
recognize that the result is universal depending only on the 
normalisation requirement $c(0)=1$ and the renormalization condition
$\s{4}{}{\vec{0};\Lam} = 1$, which together with \eq{r2},
define what we mean by the renormalized field and 
coupling $\lambda$, respectively. 

We could now argue that we should have
expected these results, without the detailed calculation. 
Nevertheless this is the first specific test of these
ideas beyond that of just cutoff function independence, and in the
process we found the
restrictions on $\hS$ sufficient to ensure scheme independence
at this level. They are merely that the seed vertices be infinitely 
differentiable and lead to convergent momentum integrals.
These conditions are needed in any case, because
the first condition ensures that no spurious infrared 
singularities are introduced \cite{Mo1,rev}, and the
second condition is necessary for the flow equation to make sense
at the quantum level and also ensures that the flow actually corresponds
to integrating out modes \cite{Mo,LM}. 

Finally, a practical method
for computing with these generalised exact RGs has been developed.
In this respect, it 
is important to stress that many of our specific choices
(what we chose to generalise in $\Psi$, how we
incorporated wavefunction renormalization, organised and solved the 
perturbative expansion) are not crucial to the calculation. 
On the contrary there are very
many ways to organise the computation; we just chose our favourite
one. The crucial step in navigating the generalised corrections, 
appears to be the recognition
that one should eliminate the elements put in by hand, in this
case vertices of $\hS$, in favour of the induced solution: the Wilsonian
effective action $S$. Indeed our computation just amounts to 
using this procedure several times over, after which many
terms are found to cancel and we are left with particularly simple 
manifestly scheme covariant results, from which we can recover 
the expected scheme independent final results.
Intuitively, this makes sense, since what are merely
our choices are
encoded in $\Psi$ (here $\hS$), whilst the actual
physics is encoded in $S$. 

For us, this is the
most important conclusion of the present paper since it implies a 
practical prescription for streamlined calculations which can be used 
even in more involved settings such as in the manifestly
gauge invariant framework
\cite{Mo1,Mo,ant}, where there is no equivalent calculation
one can directly compare to.

\section*{Acknowledgments} T.R.M. and S.A. acknowledge financial support
from PPARC Rolling grant\hfill\\ PPA/G/O/2000/00464.

\end{document}